\documentclass[prl,twocolumn,showpacs]{revtex4}%

\ifx\pdftexversion\undefined
  \usepackage[dvips]{graphics}
  \DeclareGraphicsExtensions{.eps}
\else
  \usepackage[pdftex]{graphics}
  \DeclareGraphicsExtensions{.pdf}
\fi

\usepackage{amsfonts}
\usepackage{amsmath}
\usepackage{amssymb}

\begin{document}
\title{Dephasing representation: Unified semiclassical framework for fidelity decay}
\author{Ji\v{r}\'{\i} Van\'{\i}\v{c}ek} 
\email{vanicek@post.harvard.edu}
\affiliation{Mathematical Sciences Research Institute, Berkeley, California 94720}
\affiliation{Department of Chemistry, University of California, Berkeley, California 94720}
\date{11 June 2004}

\begin{abstract}

This paper presents a unified semiclassical framework for five regimes of quantum fidelity decay and conjectures a new universal regime. The theory is based solely on the statistics of actions in the dephasing representation. Counterintuitively, in this representation, all of the decay is due to interference and none due to the decay of classical overlaps. Both rigorous and numerical support of the theory is provided.

\end{abstract}
\maketitle

While two classical trajectories with slightly different initial
conditions will exponentially diverge in a generic system, overlap of two quantum states evolved with the
same Hamiltonian $H$ remains constant for all times. The situation changes if
we consider the sensitivity to perturbations of the Hamiltonian. In a generic system, two trajectories with the same initial conditions, but propagated
by slightly different Hamiltonians $H^{0}$ and $H^{\epsilon}$ will also
exponentially diverge. The quantum-mechanical version of this question was
posed only very recently by Peres \cite{peres:1984} and has been extensively
studied in the past three years
\cite{jalabert:2001,vanicek:2003a,vanicek:2004a,jacquod:2001,cerruti:2002,prosen:2002,emerson:2002,jacquod:2003,vanicek:2003b,cucchietti:2003,kottos:2003,benenti:2003,eckhardt:2003,silvestrov:2003,yomin:2003,wang:2004,gorin:2004,bevilaqua:2004,vanicek:2004d} especially due to its relevance to quantum computation \cite{nielsen:2000}.
 The quantum-mechanical sensitivity to perturbation is measured in terms of
quantum fidelity $M\left(  t\right)  $: the overlap at time $t$ of two
initially identical states $|\psi\rangle$, that were evolved by slightly
different Hamiltonians $H^{0}$ and $H^{\epsilon}=H^{0}+\epsilon V$ ($\epsilon$ controls perturbation strength),
\[
M\left(  t\right)  =\left\vert O\left(  t\right)  \right\vert ^{2}=\vert
\langle \psi\vert e^{+i(H^{0}+\epsilon V)t/\hbar}e^{-iH^{0}t/\hbar
}\vert \psi\rangle \vert ^{2}.
\]

Recent investigations of the temporal decay of
fidelity discovered a plethora of regimes in integrable and chaotic systems.
The decay can be Gaussian
\cite{cerruti:2002,jacquod:2001}, exponential
\cite{jalabert:2001,jacquod:2001,prosen:2002} or \emph{superexponential} \cite{silvestrov:2003} in chaotic systems, and \emph{Gaussian}
\cite{prosen:2002} or \emph{algebraic} \cite{jacquod:2003} in integrable
systems. There exist actually two qualitatively different exponential regimes
in the chaotic systems: the \emph{Fermi-Golden-Rule} regime where the decay
rate is proportional to $\epsilon^{2}$ \cite{jalabert:2001,jacquod:2001}, and
the \emph{Lyapunov} regime with a perturbation-independent decay rate
\cite{jalabert:2001}. Many authors are unaware of the difference between the
\emph{perturbative} Gaussian decay in chaotic systems (which occurs only after the Heisenberg time
\cite{cerruti:2002}) and the Gaussian decay in integrable systems (which
occurs much before the Heisenberg time \cite{prosen:2002}). Generic
dynamical systems are mixed: their phase space consists of both invariant tori
and chaotic regions, and so in general fidelity decays by a mixture of the
five regimes.

Various regimes have been qualitatively explained by different quantum, statistical, or semiclassical techniques, but no single technique described more than two regimes. Moreover, semiclassical explanations always made an assumption about the decay of classical overlaps. This paper presents a unified semiclassical framework [Eq.~(\ref{fidelity_universal})] for five known regimes, using 
exclusively the statistics of actions in the dephasing representation, rendering extra assumptions unnecessary.

In Ref.\cite{vanicek:2003a}, a simple uniform (i.e., free of
singularities) semiclassical (SC) expression was derived for fidelity of initial position
states and Gaussian wave packets. In Ref.\cite{vanicek:2004a}, this expression
was rigorously justified by the shadowing theorem, and generalized to any Wigner distributions, describing both pure and mixed states.
Due to the lack of a SC dynamical prefactor, the approximation was
called the dephasing representation (DR),%
\begin{equation}
O_{DR}\left(  t\right)     =\int d^{2d}x^{\prime}\rho_{W}\left(
\mathbf{x}^{\prime}\right)  e^{i\Delta S\left(  \mathbf{x}^{\prime},t\right)
/\hbar}.
\label{wigner_dr}
\end{equation}
Here $\mathbf{x} \equiv \left(  \mathbf{r},\mathbf{p}\right)$ denotes the
phase space coordinates, $\mathbf{r}^{\prime}$ and $\mathbf{p}^{\prime}$
are the initial position and momentum, and
\begin{equation}
\Delta S\left(  \mathbf{x}^{\prime},t\right)  =S^{V}-S^{0}=-\epsilon\int
_{0}^{t}d\tau\,V\left[  \mathbf{r}\left(  \tau\right)  \right]
\label{action_difference}%
\end{equation}
is the difference at time $t$ of actions of the perturbed and unperturbed Hamiltonian
along the \emph{unperturbed }trajectory which is equal to the negative of the
integral of the perturbation along the \emph{unperturbed} trajectory. 

Although Eq.~(\ref{wigner_dr}) works for any Wigner distribution, let us consider first the most random state $\rho_W = \Omega^{-1}$, fidelity of which best approximates fidelity averaged over initial states. Here $\Omega$ stands for the phase space volume. The central result of this paper is that the five regimes occurring before the Heisenberg time can be described by a single formula,
\begin{equation}
M_{DR}\left(  t\right)  =\frac{1}{\Omega}\int d^{2d}x_{-}\exp \left [-\frac{1}{2\hbar^{2}}\left\langle\left(
\Delta S^{\prime}\mathbf{-}\Delta S^{\prime\prime}\right)  ^{2}\right\rangle  \right ],
\label{fidelity_universal}%
\end{equation}
where $\Delta S^{\prime}\equiv\Delta S\left(  \mathbf{x}^{\prime},t\right)  $,
$\Delta S^{\prime\prime}\equiv\Delta S\left(  \mathbf{x}^{\prime\prime},t\right)  $
, $\mathbf{x}_{-}\equiv\mathbf{x}%
^{\prime}-\mathbf{x}^{\prime\prime}$, and the average is over $\mathbf{x}_{+}\equiv\left(  \mathbf{x}^{\prime
}+\mathbf{x}^{\prime\prime}\right)  /2$, $\langle \cdots \rangle = \Omega^{-1} \int d^{2d} x_{+} \cdots$.
All one must do is find the variance $\langle\left(
\Delta S^{\prime}\mathbf{-}\Delta S^{\prime\prime}\right)  ^{2}\rangle $. The two criteria that yield four different regimes
are first, whether $\Delta S^{\prime}$ and $\Delta S^{\prime\prime}$ are
correlated, and second, whether the dynamics is chaotic or quasi-integrable.
Table \ref{table_variances} shows the variance in the four different cases. If
one substitutes an appropriate expression for variance from Table \ref{table_variances} into Eq. (\ref{fidelity_universal}%
) and performs the trivial integral, one obtains Table \ref{table_fidelity} of
temporal decay of fidelity in the four regimes. Table
\ref{table_regimes} associates these decays with the terminology
used in literature. 

\begin{table}
\caption{Variance $\langle(\Delta S^{\prime} - \Delta S^{\prime\prime}%
)^{2}\rangle$.}%
\label{table_variances}%
\begin{ruledtabular}
\begin{tabular}{lcc}
& \multicolumn{2}{c}{Dynamics} \\
& Chaotic &  Quasi-integrable  \\
\hline
$\Delta S^{\prime}$ and  $\Delta S^{\prime \prime }$: & & \\
\ \ Uncorrelated  & $4 K \epsilon^2 t$ & $2 C^{\infty}_{V} \epsilon^2 t^{2}$ \\
\ \ Correlated & $\frac{D}{\lambda} \alpha ^{2} \epsilon ^{2} e^{2\lambda t} x_{u}^{2}$
& $\frac{2D}{3 m^2} \epsilon ^{2} t^{3} p_{-}^{2}$ \\
\end{tabular}
\end{ruledtabular}
\end{table}

\begin{table}
\caption{Fidelity $M(t)$. $\Omega _{u}$, $\Omega _{\mathbf{p}}$ are volumes of the most unstable direction and of the momentum space; $\beta$, $\gamma$ are independent of $\epsilon$ and $t$.}
\label{table_fidelity}%
\begin{ruledtabular}
\begin{tabular}{lcc}
& \multicolumn{2}{c}{Dynamics} \\
& Chaotic &  Quasi-integrable  \\
\hline
$\Delta S^{\prime}$ and  $\Delta S^{\prime \prime }$: & & \\
\ \ Uncorrelated  & $\exp \left( -2  K \epsilon^{2} t/\hbar ^{2}\right)$ & $\exp
\left( -C^{\infty}_{V} \epsilon ^{2}  t^{2}/\hbar ^{2}\right)$ \\
\ \ Corr. (large $t$)  & $\frac{\hbar }{\alpha \Omega _{u} \epsilon } 
\sqrt{\frac{2\pi \lambda }{D}} e^{-\lambda t}$ & $\frac{1}{\Omega _{\mathbf{p}}}
\left( \frac{3\pi \hbar ^{2}m^{2}}{D\epsilon ^{2}}\right)
^{d/2}t^{-3d/2}$ \\
\ \ Corr. (small $t$)  & $ \sim \exp\left(-\beta \epsilon^2 e^{2 \lambda t}\right)$ & $ \sim e^{-\gamma \epsilon^2 t^3}$
\end{tabular}
\end{ruledtabular}
\end{table}

\begin{table}
\caption{Regimes of fidelity decay.}%
\label{table_regimes}%
\begin{ruledtabular}
\begin{tabular}{lcc}
& \multicolumn{2}{c}{Dynamics} \\
& Chaotic &  Quasi-integrable  \\
\hline
$\Delta S^{\prime}$ and  $\Delta S^{\prime \prime }$: & & \\
\ \ Uncorrelated  & Fermi-Golden-Rule & Gaussian \\
\ \ Corr. (large $t$)   & Lyapunov  & Algebraic \\
\ \ Corr. (small $t$)  & Superexponential & (Cubic-exponential) \\
\end{tabular}
\end{ruledtabular}
\end{table}

To obtain the universal expression (\ref{fidelity_universal}), one starts from the overlap (\ref{wigner_dr}) squared for random states $\rho_W=\Omega^{-1}$,
\begin{equation}
M_{DR}\left(  t\right)  =\frac{1}{\Omega^{2}} \int d^{2d}x^{\prime}\int d^{2d}%
x^{\prime\prime}\exp\left[  \frac{i}{\hbar} \left(  \Delta S^{\prime}\mathbf{-}\Delta
S^{\prime\prime}\right) \right]  .
\label{fidelity_random}
\end{equation}
Changing variables to $\mathbf{x}_{\pm}$ and averaging over $\mathbf{x}_{+}$ gives
\begin{equation}
M_{DR}\left(  t\right)  =\frac{1}{\Omega} \int d^{2d}x_{-}\left\langle \exp\left[
\frac{i}{\hbar} \left(  \Delta S^{\prime}\mathbf{-}\Delta S^{\prime\prime}\right)
\right]  \right\rangle .
\label{fidelity_random_average}
\end{equation}
If we could  isolate each regime, the difference $\Delta S^{\prime}\mathbf{-}\Delta S^{\prime\prime}$ would be
 generally Gaussian distributed (even if $\Delta S$ itself were not). Since
$\left\langle \Delta S^{\prime}\mathbf{-}\Delta S^{\prime\prime}\right\rangle
=0$, when $\Delta S^{\prime}\mathbf{-}\Delta S^{\prime\prime}$ is
Gaussian distributed, we get the universal Eq.~(\ref{fidelity_universal}).

What remains is deriving entries in Table \ref{table_variances}.
$\Delta S^{\prime}$ and $\Delta S^{\prime\prime}$ are uncorrelated if the
initial coordinates are different enough ($x_{-}$ large) or for long enough
time $t$ when the initial correlation is forgotten. Then the variance of the
difference is just twice the variance of each term, $\langle\left(  \Delta
S^{\prime}\mathbf{-}\Delta S^{\prime\prime}\right)  ^{2}\rangle=2\sigma
_{\Delta S}^{2}$, where $\Delta S$ is given by Eq.~(\ref{action_difference}).
Crucial quantity is the potential correlator $C_{V}\left(  t\right)
=\left\langle V\left[  \mathbf{r}\left(  t\right)  \right]  V\left[
\mathbf{r}\left(  0\right)  \right]  \right\rangle _{\Omega}$. In
chaotic systems, or any systems in which this correlator asymptotically decays
faster than $1/t$, $\Delta S$ follows a random walk, and 
\begin{equation}
\sigma_{\Delta S}^{2}=2 K \epsilon^{2} t
\label{variance_fgr}
\end{equation}
with  $K \equiv \int_{0}^{\infty}dt\,C_{V}\left(  t\right)  $
\cite{jalabert:2001,cerruti:2002}.
In quasi-integrable systems, or systems where $C_{V}$ asymptotically oscillates
about a finite value, $\Delta S$ follows a ballistic motion, and
\begin{equation}
\sigma_{\Delta S}^{2} = C_{V}^{\infty} \epsilon^{2} t^{2},
\label{variance_gaussian}
\end{equation}
where $C_{V}^{\infty}=\lim_{t\rightarrow\infty}t^{-1}\int_{0}^{t}d\tau
C_{V}\left(  \tau\right)  $. Analogous result was obtained quantum-mechanically by Prosen \cite{prosen:2002}.

$\Delta S^{\prime}$ and $\Delta S^{\prime\prime}$ are correlated if the
initial coordinates are close enough ($x_{-}$ small) or for short enough time
before the initial correlation is forgotten. Since $\langle\left(  \Delta
S^{\prime}\mathbf{-}\Delta S^{\prime\prime}\right)  ^{2}\rangle \neq 2\sigma
_{\Delta S}^{2}$, we cannot use a
simplification as above. Nevertheless, $\Delta S^{\prime}\mathbf{-}\Delta
S^{\prime\prime}$ itself follows a generalized random walk with a time-dependent step $\propto \delta\mathbf{r}\left( t \right) = \mathbf{r}^{\prime}(t) - \mathbf{r}^{\prime \prime}(t)$,
\begin{equation}
\Delta S^{\prime}\mathbf{-}\Delta S^{\prime\prime} \approx -\epsilon\int_{0}^{t}%
d\tau\,\nabla V\left[  \mathbf{r}\left(  \tau\right)  \right]  \cdot
\delta\mathbf{r}\left(  \tau\right),
\label{action_variation}
\end{equation}
where $\mathbf{r}(t) = (1/2)[\mathbf{r}^{\prime}(t) + \mathbf{r}^{\prime \prime}(t)]$.
If the force-force
correlator $C_{F}\left(  t\right)  =\left\langle \nabla V\left[
\mathbf{r}\left(  t\right)  \right]  \cdot\nabla V\left[  \mathbf{r}\left(
0\right)  \right]  \right\rangle _{\Omega}$ decays faster than $1/t$
then $\langle\left(  \Delta S^{\prime}\mathbf{-}\Delta S^{\prime\prime}\right)
^{2}\rangle  \approx 2 D \epsilon^{2}\int_{0}
^{t}d\tau\,\delta\mathbf{r}\left(  \tau\right)  ^{2}$ with
$D\equiv\int_{0}^{\infty}dt\,C_{F}\left(  t\right)$.
In chaotic systems, $\delta r\left(  t \right)  \approx\alpha x_{u}%
e^{\lambda t}$ where $x_{u}$ is the projection of $x_{-}$ onto the unstable
direction  and $\alpha$ depends weakly on $t$. Then
\begin{equation}
\langle\left(  \Delta S^{\prime}\mathbf{-}\Delta S^{\prime\prime}\right)
^{2}\rangle \approx D \alpha^{2}%
\epsilon^{2}e^{2\lambda t}x_{u}^{2} / \lambda.
\label{variance_lyapunov}
\end{equation}
In quasi-integrable systems, $\delta\mathbf{r}\left(  t \right)
\sim\mathbf{p}_{-}t/m$, and
\begin{equation}
\langle\left(  \Delta S^{\prime}\mathbf{-}\Delta S^{\prime\prime}\right)
^{2}\rangle  \approx 2D  \epsilon^{2}p_{-}%
^{2}t^{3} / 3m^{2}.
\label{variance_algebraic}
\end{equation}
Although it is safer to assume that $C_F(t)$ decays faster than $1/t$, this assumption seems over-restrictive because $D$ can be finite even if $C_F(t)$ decays as $1/t$ or slower, as long as it oscillates (e.g., consider $C_F(t) \sim \sin t / t$).

For long enough times, the size of phase space is irrelevant and the limits of the integral in Eq.~(\ref{fidelity_universal}) can be replaced by infinity. When the results (\ref{variance_fgr}), (\ref{variance_gaussian}), (\ref{variance_lyapunov}), (\ref{variance_algebraic}) are substituted into the universal Eq.~(\ref{fidelity_universal}), we obtain the first two rows in Table~\ref{table_fidelity} and the corresponding regimes in Table~\ref{table_regimes}. The dependence on $t$ and $\epsilon$ agrees with Refs.~\cite{jalabert:2001,jacquod:2001,cerruti:2002,prosen:2002}, \cite{prosen:2002}, \cite{jalabert:2001}, and \cite{jacquod:2003}, respectively.

For short times, when even trajectories with most distant initial conditions  are still correlated, size of phase space comes into play, and substituting variances (\ref{variance_lyapunov}) and (\ref{variance_algebraic}) into Eq.~(\ref{fidelity_universal}) gives the last row of Table~\ref{table_fidelity} and Table~\ref{table_regimes}.
The first result agrees with Ref.~\cite{silvestrov:2003}, the latter predicts a short-time cubic-exponential decay in quasi-integrable systems, yet to be observed. 

While the DR (\ref{wigner_dr}) has already been tested for position eigenstates in chaotic \cite{vanicek:2003a,vanicek:2003b,wang:2004} and mixed \cite{vanicek:2004a} systems, detailed numerical tests of the more general expression (\ref{wigner_dr}) for both pure and mixed states, in chaotic, mixed, and
quasi-integrable systems, in the first five regimes from Table \ref{table_regimes} as well as intermediate regimes will be presented elsewhere \cite{vanicek:2004d}.
Here the numerics is focused on the verification of Table~\ref{table_variances} which is the only input required by Eq.~(\ref{fidelity_universal}). In particular, the dependence of $\langle\left(
\Delta S^{\prime}\mathbf{-}\Delta S^{\prime\prime}\right)  ^{2}\rangle$ on $p_{-}$ and $t$ is checked. Many authors believe that it is necessary to calculate fidelity averaged over initial states to obtain regimes in the second row of Table~\ref{table_regimes}. To disprove that belief let us use pure position states $\rho_W = \Omega^{-1}_{\mathbf{p}} \delta (\mathbf{r} - \mathbf{R})$. Derivations in Eqs.~(\ref{fidelity_universal})-(\ref{fidelity_random_average}) will stay the same  if we replace integrals $\Omega ^{-1} \int d^{2d}x \cdots $ by $\Omega ^{-1}_{\mathbf{p}} \int d^{d} p \cdots$.

For numerics, a perturbed standard map \cite{vanicek:2004a},
\begin{alignat*}{2}
p_{n+1} &= p_{n} + k \sin {q}_n + \epsilon \sin 2 q_{n} \ \ \ & &(\mathrm{mod}\, 2\pi), \\
q_{n+1} &= q_{n} + p_{n + 1}  & &(\mathrm{mod}\, 2\pi) .
\end{alignat*}
was used. Here $q_n$, $p_n$ are the position and momentum at discrete times $n$, and $k$ is a parameter controlling the transition from integrability to chaos. In the calculations described below, $k=20$ was used as a representative chaotic system, and $k=0.3$ as a representative quasi-integrable system. The initial state is  a position state $Q = 0.8 \, \pi$, avoiding any problems due to symmetry. Between 500 and 1000 trajectories were used to check the statistics of actions. Effective Planck constant is $\hbar = 1 / 2\pi n$ where $n$ is the size of Hilbert space. Specifically, $n=1000$ and $\epsilon = 0.003$ were used for the chaotic example, and
$n = 100$ and $\epsilon = 0.005$ for the quasi-integrable case. 

\begin{figure}
\centerline{\resizebox{\hsize}{!}{\includegraphics{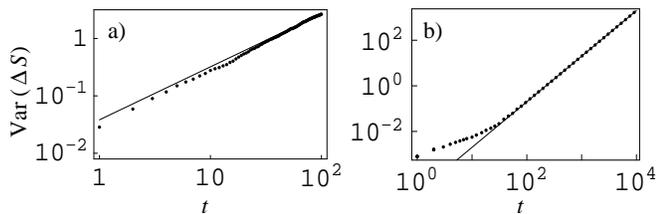}}}\caption{Variance
of $\Delta S$ as a function of time in a a) chaotic  and b) quasi-integrable 
system. Dependence is a) linear (slope $\approx$ 0.992) and b) quadratic (slope $\approx$ 1.9995).}%
\label{fig_act_var_dep_time}
\end{figure}

\begin{figure}
\centerline{\resizebox{\hsize}{!}{\includegraphics{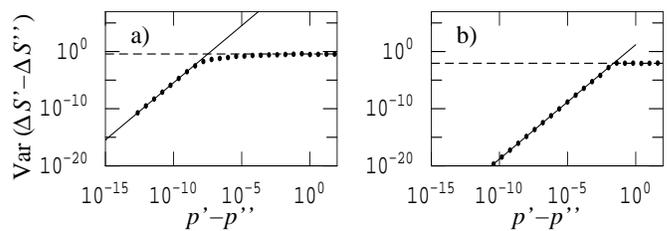}}}
\caption{Variance of $\Delta S^{\prime} - \Delta S^{\prime\prime}$ at a fixed time $t=7$ as a function of $p_{-} = p^{\prime} - p^{\prime\prime}$ in a a) chaotic and b)
quasi-integrable system. In both cases the dependence is first quadratic and then independent of $p^{\prime} - p^{\prime\prime}$. Fitted slopes: a) 1.998 and 0.01, b) 2.000 and 0.001.}
\label{fig_var_dep_mom}
\end{figure}

\begin{figure}
\centerline{\resizebox{\hsize}{!}{\includegraphics{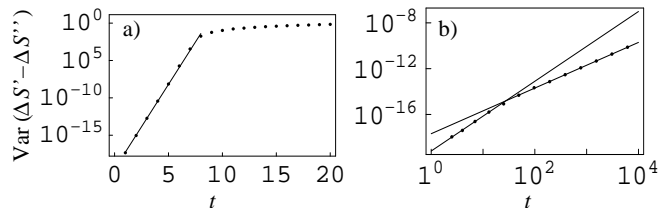}}}\caption{Variance
of $\Delta S^{\prime} - \Delta S^{\prime\prime}$ as a function of time for fixed $p^{\prime} - p^{\prime\prime} = 10^{-9}$ in a a)
chaotic  and b) quasi-integrable system. The dependence is: a) first  exponential, then linear, and b) first cubic (slope $\approx$ 3.046), then quadratic (slope $\approx$ 1.994). In b), dots represent the numerical variance averaged over a time interval $(t/2,t)$ since the variance itself has large oscillations about the averaged value.}%
\label{fig_diff_act_var_dep_time}%
\end{figure}

Dependence of $\sigma_{\Delta S}^{2}$ on time, described by Eqs.~(\ref{variance_fgr}) and (\ref{variance_gaussian}), is verified in Fig.~\ref{fig_act_var_dep_time}. While part a) shows that in chaotic systems, $\sigma_{\Delta S}^{2}$ grows linearly with time, part b) shows that in quasi-integrable systems, $\sigma_{\Delta S}^{2}$ grows quadratically with time.
Fig.~\ref{fig_var_dep_mom} verifies the dependence of $\langle\left(  \Delta S^{\prime} - \Delta S^{\prime\prime}\right)^2 \rangle$ at a fixed time on the difference $p_{-}$ of initial momenta. Parts a) and b) show that in both chaotic and quasi-integrable systems, this dependence is quadratic for small $p_{-}$ (as in Eqs.~(\ref{variance_lyapunov}) and (\ref{variance_algebraic})) and independent of $p_{-}$ for large $p_{-}$ (as in Eqs.~(\ref{variance_fgr}) and (\ref{variance_gaussian})).  The transition occurs for  $p_{-}$ such that by the time $t$, two trajectories with initial distance $p_{-}$ completely lose their correlation.
Dependence of $\langle\left(  \Delta S^{\prime} - \Delta S^{\prime\prime}\right)^2 \rangle$ on time for a given difference of initial momenta is confirmed in Fig.~\ref{fig_diff_act_var_dep_time}. Part a) shows that in chaotic systems this dependence is first exponential, as in Eq.~(\ref{variance_lyapunov}), and later linear, as in Eq.~(\ref{variance_fgr}) (hard to see here, but can be seen in a log-log plot such as in Fig.~\ref{fig_act_var_dep_time}). Part b) shows that in quasi-integrable systems, this dependence is first cubic, as in Eq.~(\ref{variance_algebraic}), and then quadratic, as in Eq.~(\ref{variance_gaussian}). The transition occurs at time $t$ when two trajectories with initial distance $p_{-}$ completely lose their correlation. 

Intuitively, one would think that the decay of quantum fidelity would have two
components: the decay of classical overlaps (classical fidelity) and the decay
due to dephasing (destructive interference). This approach was taken in all
previous fidelity literature. Contrary to that,
in the DR, all of the decay is due to interference. This
can be best seen in the algebraic decay: while in
Ref.\cite{jacquod:2003}, the decay of classical overlaps, $\propto t^{-d}$,
and the decay due to dephasing,
$\propto t^{-d/2}$, together give the overall fidelity decay, $\propto
t^{-3d/2}$, in the present approach, the same overall decay $t^{-3d/2}$ is entirely due to
dephasing in DR.

I would like to point out the rigorousness of going from $\langle\left(
\Delta S^{\prime}\mathbf{-}\Delta S^{\prime\prime}\right)  ^{2}\rangle$ to
$M\left(  t\right)  $ in the DR approach: here the
integration is done analytically and correctly. Similar approach based on the
statistics of action differences (however, in the \emph{final position}
representation where the interference accounts only for a part of the decay) was used in literature to derive the Lyapunov
\cite{jalabert:2001} and algebraic \cite{jacquod:2003} decay. Both Refs.~\cite{jalabert:2001} and
\cite{jacquod:2003} provide a long derivation in which there appears an
integral of the form
\[
\int d^{d}\mathbf{r}\sum_{j}\left\vert \det({\partial\mathbf{p} _{j}^{\prime}%
}/{\partial\mathbf{r}})\right\vert ^{2}\cdots,
\]
in particular, there is a second power of a certain Jacobian (Van Vleck
determinant). In both Refs.~\cite{jalabert:2001,jacquod:2003}, a suspicious change of variables is employed, in which
one power of the Jacobian is used to change variables from $\mathbf{r}$ to $
\mathbf{p}^{\prime}$ and the other power of the Jacobian is replaced by its
estimate to get
\[
\int d^{d}\mathbf{p}^{\prime}\left\vert \det({\partial\mathbf{p} ^{\prime}%
}/{\partial\mathbf{r}})\right\vert _{\text{estimate}}\cdots.
\]
While both Refs. obtain the right overall behavior at the end (because they guess the behavior of the Jacobian), this
step makes the lengthy mathematical derivation non-rigorous.

Originally, dephasing representation (\ref{wigner_dr}) was used only in chaotic systems
\cite{vanicek:2003a}. Besides generalizing DR to arbitrary Wigner functions, the main result of Ref.~\cite{vanicek:2004a} was
showing that DR is valid for finite times also in mixed and
quasi-integrable systems. The mathematical cornerstone of the approximation is
the shadowing theorem \cite{hammel:1987,grebogi:1990,chow:1992,coomes:1994,hayes:2003} which guarantees the existence of an
unperturbed trajectory with slightly different initial conditions that is near
(shadows) a perturbed trajectory up to time $t_s$. In uniformly hyperbolic
systems, time $t_s$ is infinite, but for a much larger class of systems,
shadowing works for finite times $t_s$. In its original form \cite{hammel:1987,grebogi:1990,chow:1992,coomes:1994,hayes:2003}, shadowing theorem
justifies  computer-generated trajectories in chaotic systems
 where perturbation is a random noise due to
the round-off errors. The relation with deterministic Hamiltonian
perturbations was rigorously shown in Ref.\cite{vanicek:2004a}.

 Dephasing representation (\ref{wigner_dr}) appears too simple to be true because Eq.~(\ref{action_difference}) has the form of first order perturbation approximation and appears to consider only the interference of the ``diagonal'' terms. The DR seems to require that the change of the trajectory due to perturbation be very small. While some authors believe that, this is not true at all.  Other authors believe that the apparent first-order perturbation approximation is  a convenience and that it could be improved by considering the exact trajectories. That is not true either.  The reason why DR works is the shadowing theorem as explained in Ref.\cite{vanicek:2004a}. While a perturbed trajectory with the same initial condition can change enormously, there exists a perturbed trajectory with slightly different initial condition that changes very little. Trying to improve the approximation by relieving either apparent ingredient (``diagonal approximation'' or ``first order perturbation approximation'') not only does not improve the numerical results, but actually gives much worse results.

To conclude, a simple unified framework was presented for the temporal decay of quantum fidelity. It is based solely on the statistics of actions in the DR, through Eq.~(\ref{fidelity_universal}), and can describe five known and one new universal regime. The DR (\ref{wigner_dr}) \cite{vanicek:2003a,vanicek:2004a} is even more general and can represent quantum fidelity in a mixture of regimes as well as in other non-universal regimes \cite{wang:2004}.  

The author wishes to thank the Mathematical Sciences Research Institute for financial support.

\end{document}